\documentclass[11pt,twoside]{article}

\usepackage{asp2006}
\usepackage{epsf}
\usepackage{psfig}
\usepackage{lscape}
\usepackage{graphicx}

\begin{document}
\title{Three Cycles of the Solar Toroidal Magnetic Field and This Peculiar Minimum}   %%% Fill in title
\author{LO, Leyan, HOEKSEMA, J.T. and SCHERRER, P.H.}   %%% Fill in author names
\affil{HEPL, Stanford University, Stanford, CA 94305-4085, USA}    %%% Fill in author affiliations

\begin{abstract} %%% Abstract to run on from here.
Thirty-four years of WSO (Wilcox Solar Observatory) and thirteen years of SOHO/MDI (Michelson Doppler Imager
on the Solar and Heliospheric Observatory) 
magnetograms have been studied to measure the east-west inclination angle, indicating the toroidal component 
of the photospheric magnetic field. This analysis reveals that the large-scale toroidal component of the 
global magnetic field is antisymmetric around the equator and reverses direction in regions associated with 
flux from one solar cycle compared to the next. The toroidal field revealed the first early signs of cycle 24 
at high latitudes, especially in the northern hemisphere, appearing as far back as 2003 in the WSO data and 
2004 in MDI. As in previous cycles, the feature moves gradually equatorward. Cycles overlap and 
the pattern associated with each cycle lasts about 17 years. Even though the polar field at the current 
solar minimum is significantly lower than the three previous minima, the toroidal field pattern is 
similar.
\end{abstract}

\section{Introduction}
Solar dynamo models predict that the toroidal and poloidal components of the global magnetic field are 
regenerated from one other. The poloidal field is transformed into the toroidal field from differential 
rotation (the $\Omega$-effect), and the toroidal field is twisted into the poloidal field (the $\alpha$-effect). 
These alternate and repeat in a 22 year cycle [e.g. \citeauthor{dg2008}, 2008].

This toroidal field component, previously measured by \citeauthor{ss1994} [1994] using WSO data over the 
period of 1977-1992, provided evidence for an extended activity cycle of 16-18 years. 
\citeauthor{ub2005} [2005] also measured the toroidal field using Mount Wilson data from 1986-2004 and 
verified the dynamo model for the creation and reversal of the toroidal field.

Since we are facing a peculiar solar minimum, we want to investigate whether there is also some peculiar 
activity in the toroidal fields leading up to the minimum. Three solar cycles of WSO magnetic field data are available 
to us now, as well as a complete cycle observed with SOHO/MDI [\citeauthor{sea1995}, 1995].

\section{Method \& Results}
We first generated maps of the inclination angle for both the positive and negative field polarities using 
the WSO magnetograms and the MDI synoptic maps. This is done by tracking the line-of-sight magnetic field,
$B_l$, for a given Carrington coordinate as it crosses the disk.
\begin{figure}[!ht]
\center
\includegraphics[height=144pt]{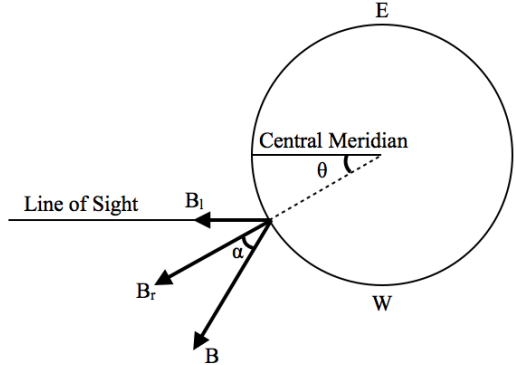}
\caption{Diagram of the field components and angles as viewed from the north pole. The east-west field 
inclination angle, $\alpha$, is the angle between $B$ and $B_r$. We measure $B_l$ as it crosses the disk 
over various $\theta$'s.}\label{components}
\end{figure}
We can fit the measurements of $B_l$ for different longitudes to the following equation to calculate 
the average east-west inclination angle of the field, $\alpha$:
\begin{equation}
B_l(\theta) = A \cos(\theta + \alpha)
\end{equation}
Where $A$ is proportional to the true magnetic field. A derivation of this fit gives the following 
formula (\citeauthor{ss1994}, 1994):
\begin{equation}
\tan \alpha = \frac{(\displaystyle\sum_i B_l^i \cos \theta_i)(\displaystyle\sum_i \sin \theta_i \cos \theta_i) - 
(\displaystyle\sum_i B_l^i \sin \theta_i)(\displaystyle\sum_i \cos^2 \theta_i)}{(\displaystyle\sum_i B_l^i \cos \theta_i)
(\displaystyle\sum_i \sin^2 \theta_i) - (\displaystyle\sum_i B_l^i \sin \theta_i)(\displaystyle\sum_i \sin \theta_i \cos \theta_i)}
\end{equation}
We fit the values of $B_l$ and $\theta$ into the equation for $\alpha$ to generate maps of the 
inclination angles. 
Following \citeauthor{ss1994}, individual regions are tracked across the WSO magnetograms.
For MDI, measurements of the same features in synoptic maps constructed
using data from central meridian, $\pm15^\circ, \pm30^\circ$ and $\pm45^\circ$ are compared.
We generate two separate maps for each Carrington rotation,
one for positive polarity $B_l$ and the other for negative. 
The difference of these two maps gives the inclination difference. 
This inclination difference map gives us the east-west direction of the magnetic field at 
each point on the sun for each Carrington rotation.

\begin{figure}[!ht]
\center
\plotone{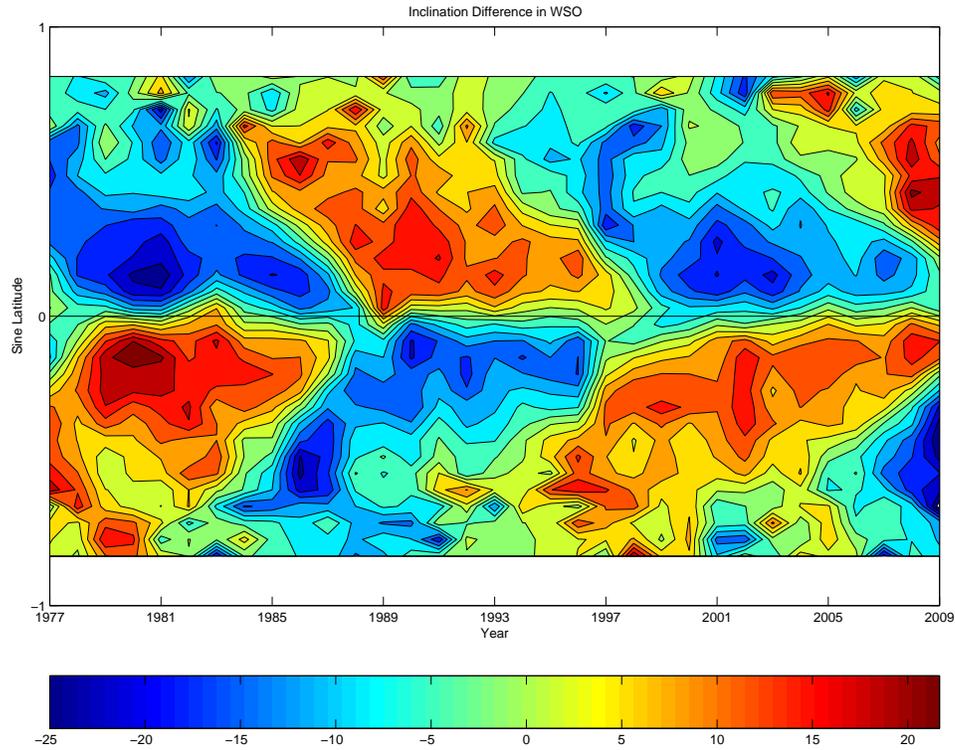}
\caption{Contour plot of the rotation averages of inclination differences
versus latitude from 1977-2009 using WSO data.  There are almost three
complete cycles of inclination differences, and the extended solar
minimum can be seen in both the duration of cycle 23 near the equator,
as well as in the rate at which cycle 24 moves equatorward. Toroidal
field assiciated with Cycle 24 first emerges in 2003 in the north at
high latitudes and the pattern looks almost identical to the last two
cycles.}
\label{incdiff_wso_all} 
\end{figure}

\begin{figure}[!ht]
\center
\includegraphics[height=150pt]{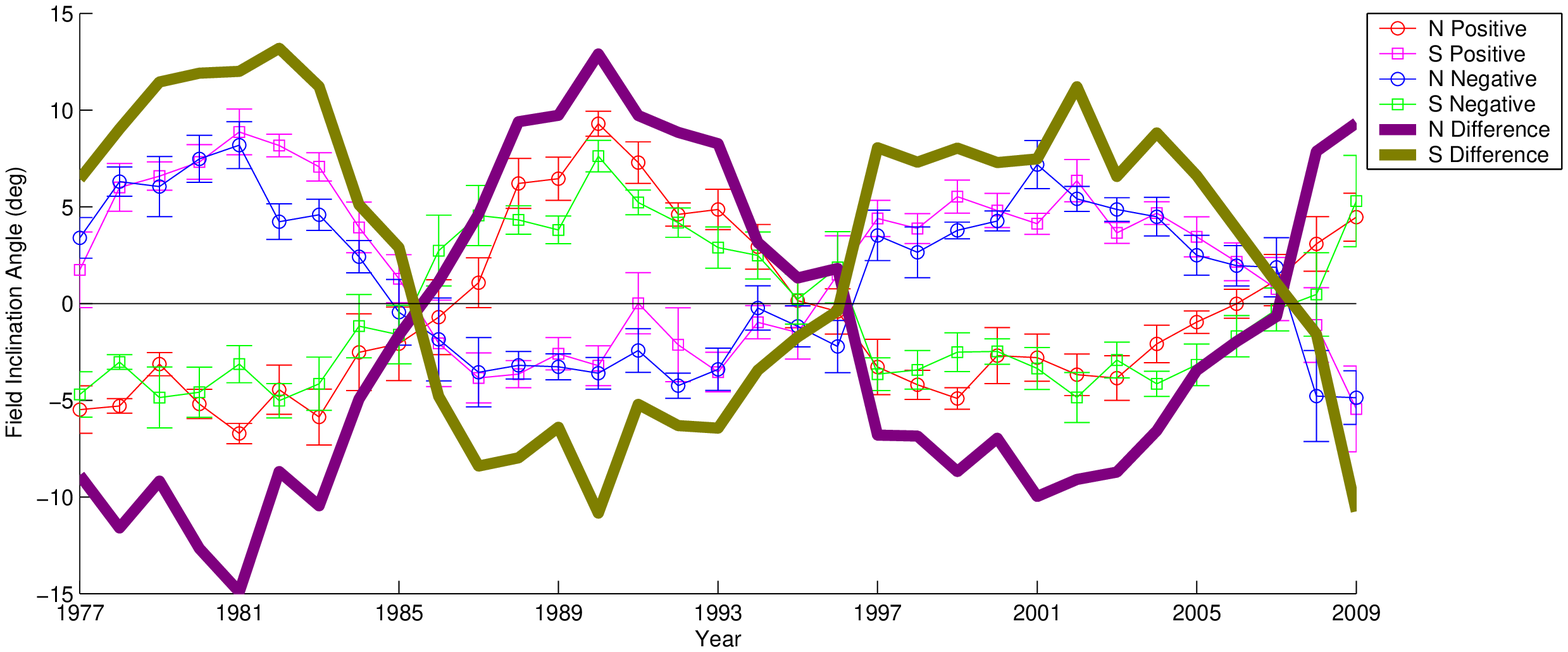}
\caption{Line plot of the summary averages of WSO inclinations for each hemisphere from 1977-2009.
The lighter lines show each polarity separately.  The heavy red (green) line shows the inclination difference in the north (south).
In 2007, the average northern and southern 
field angles flip sign, indicating the start of Cycle 24. The sign changes when most of the weak 
field regions in each hemisphere have reversed their toroidal field components. Oddly enough, the extended 
minimum is not observed in the line plot; the interval between zero crossings this cycle is the same as 
last.}\label{incangle_wso_all}
\end{figure}

The inclination difference maps for each rotation are averaged over all longitudes
to show the inclination differences at each latitude over time, 
using both WSO and MDI data. From these averages 
we construct the contour plots shown in Figure \ref{incdiff_wso_all}, 
and the line plots (Figure \ref{incangle_wso_all}) for each hemisphere. 
The WSO data span 34 years, which gives us three solar cycles of coverage.

\begin{figure}[!ht]
%%% Figure labels are too small to read
\plottwo{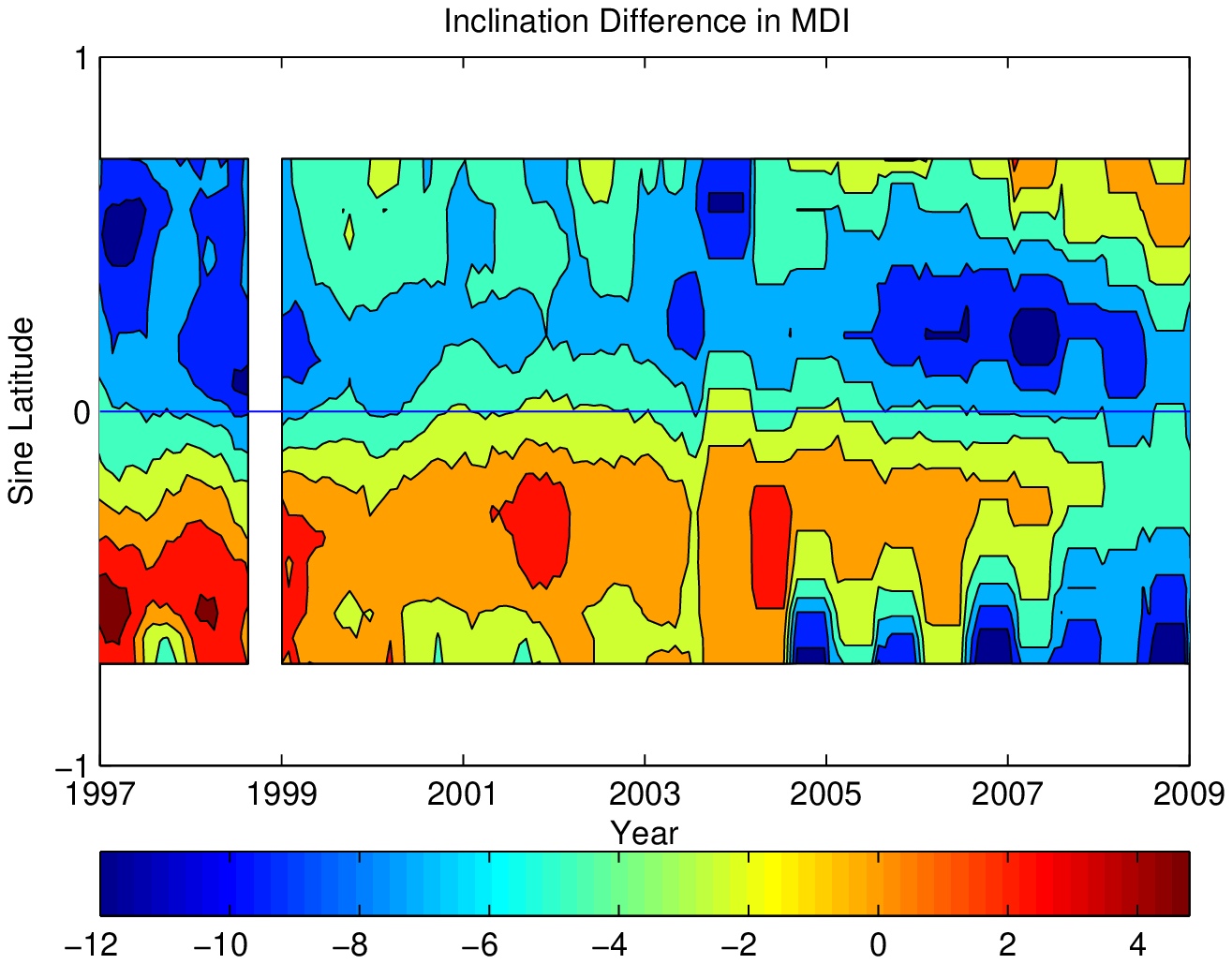}{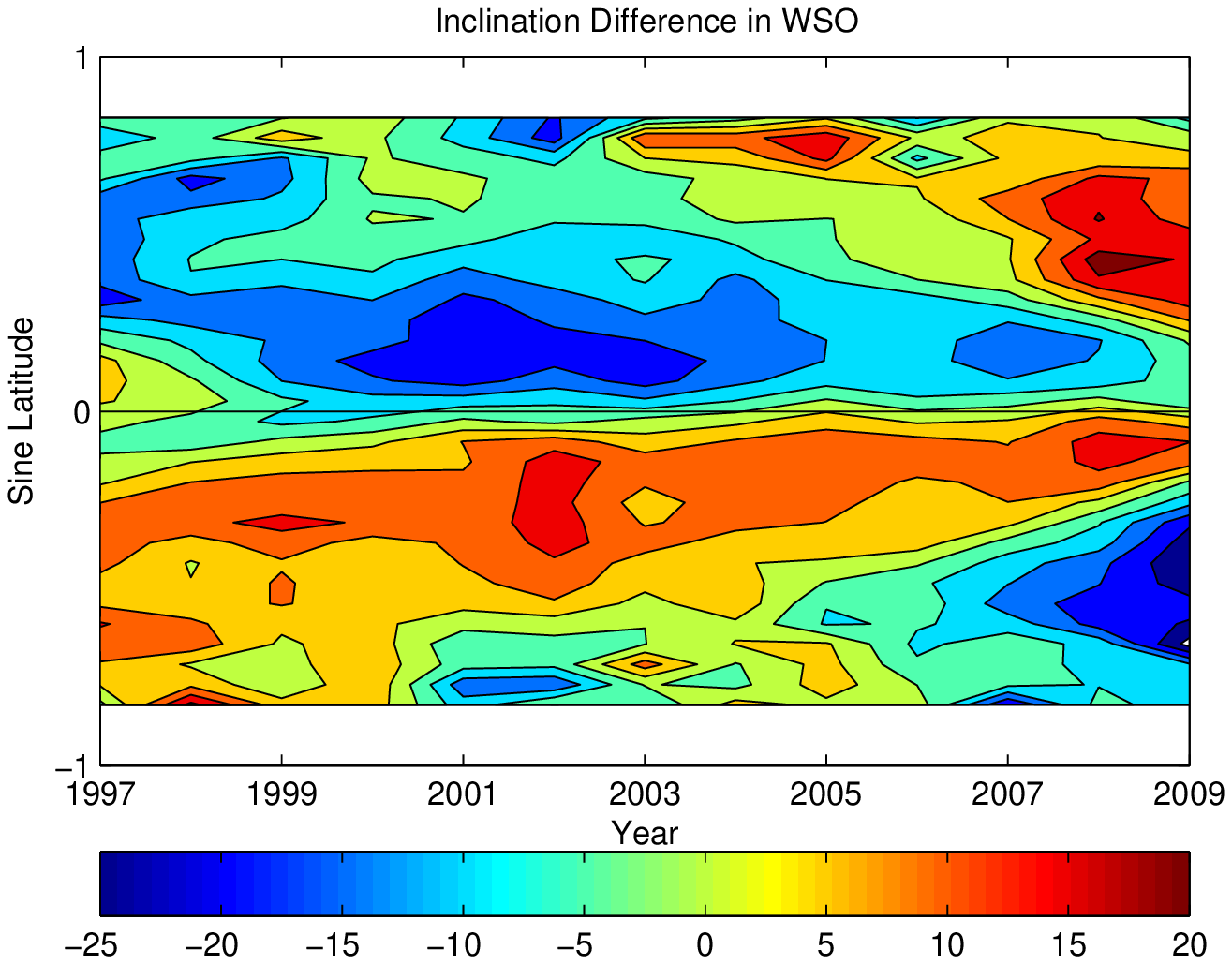}
\caption{The plot generated from MDI (left) span only one cycle, from 1997-2009. 
The general shape of the contour map matches WSO's for this time period (right). 
Both MDI and WSO show an extended Cycle 23 and the emergence of Cycle 24 in 2004 
in the north at high latitudes.  Periodic rolls of the spacecraft after 2003 add 
noise to MDI.}
\label{incdiff_compare}
\end{figure}

However, the MDI values are somewhat different than WSO, as can be seen in 
Figure \ref{incdiff_compare}. We believe this is due to asymmetric noise characteristics of the MDI instrument. 
In 2003, SOHO began to periodically roll from $0^\circ$ to $180^\circ$, which caused periodic 
irregularities in our data that need to be better calibrated.

\section{Summary}
From the inclination difference summary averages, we detected the first emergence of Cycle 24 
some time during 2003-2004. This emergence pattern is very much like the pattern observed in 
the previous two cycles. The new emerging inclination angles, which are related to the toroidal 
fields, are consistent with those from the previous cycles.  This suggests that the toroidal field 
is not a clear indicator of the duration of the following solar minimum or the generation of the 
poloidal fields at the concurrent minimum.  However, if the toroidal field is generated from the 
preceding poloidal field, we might expect to see that the next cycle's toroidal field will be weak. 
Also, the inclination angle that indicates the toroidal field at the Sun's surface is only a 
distant reflection of the toroidal field that is present wherever the true solar dynamo is acting. 
Particularly in the higher resolution MDI data, the weak field regions that contribute to the 
analysis may also reflect the spatially averaged characteristics of a surface dynamo process whose 
contribution to the solar cycle dynamo is uncertain.
Further work needs to be done to solve the calibration issues of MDI.

\acknowledgements %%% Text of acknowledgements runs on after this command.

This work was supported by NASA under MDI Grant NNX09AI90G and WSO Grant NNX08AG47G.  
SOHO is a project of international cooperation between ESA and NASA.

%%% THE BIBLIOGRAPHY
%%%
%%% CONSULT SECTION 3 OF "INSTRUCTIONS FOR AUTHORS" FOR HOW TO USE NATBIB.
%%% AUTHORS ARE ENCOURAGED TO USE EITHER THE "THEBIBLIOGRAPY" ENVIRONMENT
%%% BY UNCOMMENTING (DELETING THE "%" SYMBOL) THE COMMANDS BELOW, OR BY
%%% USING THE BIBTEX ENVIRONMENT. TO FIND OUT WHICH IS APPLICABLE TO YOUR
%%% CONTRIBUTION, CONSULT THE VOLUME EDITORS FOR YOUR PROCEEDINGS.
%%%

\end{document}